\documentclass[3p,twocolumn]{elsarticle}

\usepackage{lineno}
\usepackage{hyperref}
\usepackage{graphicx}
\usepackage{multirow}
\usepackage[fleqn]{amsmath}
\usepackage[usenames,dvipsnames]{xcolor}

\modulolinenumbers[5]

\journal{Molecular Astrophysics}

\usepackage{numcompress}

\begin{document}

\begin{frontmatter}

\title{High-resolution absorption cross sections of C$_{2}$H$_{6}$ at elevated temperatures}

\author[ODUaddress]{Robert J. Hargreaves\corref{mycorrespondingauthor}}
\cortext[mycorrespondingauthor]{Corresponding author}
\ead{rhargrea@odu.edu}

\author[ODUaddress]{Eric Buzan}

\author[ODUaddress]{Michael Dulick}

\author[ODUaddress]{Peter F. Bernath}

\address[ODUaddress]{Department of Chemistry, Old Dominion University, 4541 Hampton Boulevard, Norfolk, VA 23529, USA}

\begin{abstract}
Infrared absorption cross sections near 3.3 $\mu$m have been obtained for ethane, C$_{2}$H$_{6}$. These were acquired at elevated temperatures (up to 773 K) using a Fourier transform infrared spectrometer and tube furnace with a resolution of 0.005 cm$^{-1}$. The integrated absorption was calibrated using composite infrared spectra taken from the Pacific Northwest National Laboratory (PNNL). These new measurements are the first high-resolution infrared C$_{2}$H$_{6}$ cross sections at elevated temperatures.
\end{abstract}


\end{frontmatter}


\section{Introduction}
\label{sect1}

Ethane (C$_{2}$H$_{6}$) is the second largest component of natural gas and is primarily used in the industrial manufacture of petrochemicals. It is present as a trace gas in the Earth's atmosphere and can be used to monitor anthropogenic (e.g., fossil fuel emission, combustion processes) and biogenic sources \cite{2011Natur.476..198A, 2007JQSRT.107..407K, 2011ACP....1112169T}.

However, C$_{2}$H$_{6}$ is also of particular interest to astronomy. C$_{2}$H$_{6}$ is found in all four giant planets \cite{1974ApJ...187L..41R, 1981Sci...212..192H, 1987Icar...70....1O}, Titan \cite{2005Natur.438..779N}, comets \cite{1996Sci...272.1310M} and even as an ice in Kuiper Belt objects \cite{2007AJ....133..284B}. For Titan, observations indicate C$_{2}$H$_{6}$ is a constituent of light hydrocarbon lakes \cite{2008Natur.454..607B}.  In the atmospheres of the giant planets and Titan, C$_{2}$H$_{6}$ is primarily formed from the photolysis of methane, CH$_{4}$ \cite{2007Icar..188...47N, 2008SSRv..139..191M}, and subsequent recombination of methyl radicals, CH$_{3}$ \cite{2009JPCA..11311221W, 2009Icar..201..226K, 2014Icar..236...83K}.

In Jupiter, stratospheric observations have detected hot C$_{2}$H$_{6}$ in polar auroral regions \cite{2009Icar..202..354K}. These hot spots occur close to similar hot CH$_{4}$ and H$^{+}_{3}$ emission \cite{2015Icar..257..217K} and are heated due to the channeling of particles by the strong magnetic field. The Juno mission \cite{2007AcAau..61..932M} is due to arrive at Jupiter in 2016 and one major objective for the Jovian Infrared Auroral Mapper, JIRAM \cite{2008AsBio...8..613A}, is to study these auroral hot spots to determine the molecules responsible and their vertical structure.

Brown dwarfs are sub-stellar objects that do not burn hydrogen in their cores \cite{2005ARA&A..43..195K}. They are warm (albeit relatively cool in comparison to stars), thereby allowing for the formation of rich molecular atmospheres. Brown dwarf atmospheric chemical models predict C$_{2}$H$_{6}$ to form deep in these objects \cite{2002Icar..155..393L}. Recent observations indicate that these objects can also harbour extremely bright aurorae \cite{2015aurora}. Similarly, exoplanets known as hot-Jupiters orbit close to their parent star and have atmospheric temperatures conducive to molecule formation. While models predict C$_{2}$H$_{6}$ may have a low thermochemical abundance in the atmosphere of these objects \cite{2012A&A...546A..43V}, disequilibrium and increased metallicity can lead to significant enhancements \cite{2011ApJ...738...32L, 2011ApJ...737...15M, 2013MNRAS.435.1888B}. C$_{2}$H$_{6}$ can be used as a useful temperature probe for exoplanets and brown dwarfs \cite{2013A&ARv..21...63T}, but high temperature laboratory data are missing. It is therefore important to have high temperature data available for astronomical and terrestrial applications.

Due to the prevalence of C$_{2}$H$_{6}$, the infrared spectrum has been the focus of numerous studies, but complete line assignments are difficult to obtain. This is, in part, due to the $\nu_{4}$ torsional mode near 35 $\mu$m (290 cm$^{-1}$) \cite{1999JChPh.111.9609M, 2001JMoSp.209..228M, 2008JMoSp.250...51B, 2015JQSRT.151..123M} which produces numerous low frequency hot bands, extensive perturbations and a very dense line structure. The $\nu_{9}$ mode near 12 $\mu$m (830 cm$^{-1}$), often used in remote sensing \cite{2007ACP.....7.5861C, 2015Icar..250...95V}, has been the focus of comprehensive analyses that have significantly improved line assignments \cite{2007ApJ...662..750V, 2010JQSRT.111.1234M, 2010JQSRT.111.2481M}. Line parameters and assignments have been obtained for the $\nu_{8}$ band near 6.8 $\mu$m (1470 cm$^{-1}$) \cite{2008JMoSp.248..134L, 2011MolPh.109.2219L, 2012PandSS...60...93D} as well as the $\nu_{5}$ and $\nu_{7}$ modes contained in the 3.3 $\mu$m (3000 cm$^{-1}$) spectral region \cite{2011JGRE..116.8012V, 2011JMoSp.267...71L}. Although considerable progress has been made in these recent studies, high-resolution analyses are generally incomplete and still fail to match laboratory observations precisely.

The Pacific Northwest National Laboratory (PNNL) has recorded infrared absorption cross sections for a large number of species, including C$_{2}$H$_{6}$ (see \href{http://nwir.pnl.gov}{http://nwir.pnl.gov} and Ref. \cite{2004ApSpe..58.1452S}), at three temperatures (278, 293 and 323 K). High-resolution (0.004 cm$^{-1}$) absorption cross sections have been provided at room temperature \cite{2010JQSRT.111..357H} and these measurements constitute the C$_{2}$H$_{6}$ cross sections contained in HITRAN \cite{2013JQSRT.130....4R}. However, the intended use of these data are for the study of the Earth's atmosphere and will give an incorrect radiative transfer when applied to high temperature environments. High-temperature absorption cross sections of hydrocarbon species (including C$_{2}$H$_{6}$) have been recorded for combustion applications \cite{2014JMoSp.303....8A} {\color{Black}at relatively low} resolution ($\ge$0.16 cm$^{-1}$) {\color{Black}and are} not sufficient for high-resolution applications. 

The aim of this work is to provide high-resolution absorption cross sections of C$_{2}$H$_{6}$ at elevated temperatures to be used in the analysis of brown dwarfs, exoplanets and auroral hot spots of Jupiter.

\section{Measurements}
\label{sect2}

Spectra were acquired of C$_{2}$H$_{6}$ between 2200 and 5600 cm$^{-1}$ (1.8 $-$ 4.5 $\mu$m) using a Fourier transform infrared spectrometer at a resolution of 0.005 cm$^{-1}$. These spectra cover the temperatures 296 $-$ 773 K and experimental conditions are provided in Table~\ref{tab1}.

\begin{table*}[t!]
  \caption{Experimental conditions and Fourier transform parameters}
  \label{tab1}
  \centering
  \begin{small}
  \begin{tabular}{@{}lc@{}}
  \rule{0pt}{1ex}  \\
  \hline
  \rule{0pt}{1ex}  \\
   Parameter    &  Value\textit{$^{a}$}  \\
  \rule{0pt}{1ex}  \\
 \hline
  \rule{0pt}{1ex}  \\
Temperature range                     & 296 $-$ 773 K \\
Spectral range                        & 2200 $-$ 5600 cm$^{-1}$ \\
Resolution                            & 0.005 cm$^{-1}$ \\
Path length                           & 0.5 m \\
Sample cell material                  & Quartz (SiO$_{2}$) \\
External source                       & External globar\textit{$^{b}$} \\
Detector                              & Indium antimonide (InSb) \\
Beam splitter                         & Calcium fluoride (CaF$_{2}$) \\
Spectrometer windows                  & CaF$_{2}$ \\
Filter                                & Germanium  \\
Aperture                              & 1.5 mm \\
Apodization function                  & Norton-Beer, weak \\
Phase correction                      & Mertz \\
Zero-fill factor                      & $\times$16 \\
  \rule{0pt}{1ex}  \\
\hline
  \rule{0pt}{1ex}  \\
\multicolumn{2}{p{8cm}}{\textit{$^{a}$} all spectra recorded under same conditions except where stated. } \\
\multicolumn{2}{p{8cm}}{\textit{$^{b}$} no external source for $B_{\textrm{\scriptsize{em}}}$ and $B_{\textrm{\scriptsize{ref}}}$. } \\
\end{tabular}
\end{small}
\end{table*}

The spectrometer is combined with a tube furnace containing a sample cell made entirely from quartz, thereby allowing the cell to be contained completely within the heated portion of the furnace \cite{2015ApJ..inpressH}. At elevated temperatures, the C$_{2}$H$_{6}$ infrared spectrum has both emission and absorption components. The emission components {\color{Black}can be} included in the final transmittance spectra by following the same procedure outlined in Ref. \cite{2015ApJ..inpressH} for CH$_{4}$. This involves recording both C$_{2}$H$_{6}$ absorption ($A_{\scriptsize{\textrm{ab}}}$) and emission ($B_{\scriptsize{\textrm{em}}}$) spectra, then combining as
\begin{equation}
\label{eqn1}
\tau = \frac{A_{\scriptsize{\textrm{ab}}} -– B_{\scriptsize{\textrm{em}}}}{A_{\scriptsize{\textrm{ref}}} –- B_{\scriptsize{\textrm{ref}}}},
\end{equation}
to give the transmittance spectrum ($\tau$), where $A_{\scriptsize{\textrm{ref}}}$ and $B_{\scriptsize{\textrm{ref}}}$ are the background reference spectra of the absorption and emission, respectively. {\color{Black}The emission component of C$_{2}$H$_{6}$ is sufficiently strong at 673 and 773 K that an emission correction is necessary; therefore $B_{\scriptsize{\textrm{em}}}$ and $B_{\scriptsize{\textrm{ref}}}$ are required. For lower temperatures, $B_{\scriptsize{\textrm{em}}}$ and $B_{\scriptsize{\textrm{ref}}}$ equal zero and Equation~\ref{eqn1} reverts to the standard transmittance equation (i.e., $\tau = A_{\scriptsize{\textrm{ab}}}/A_{\scriptsize{\textrm{ref}}}$).}

The C$_{2}$H$_{6}$ infrared spectrum near 3000 cm$^{-1}$ (3.3 $\mu$m) contains a small number of $\nu_{7}$ $Q$-branch features that are significantly stronger than the $P$- and $R$-branches and the nearby $\nu_{5}$ mode.  In order to maximize the signal from the weaker features, the C$_{2}$H$_{6}$ spectra were acquired at both ``high'' and ``low'' pressure. These high and low pressure experiments are summarised in Table~\ref{tab2}. The low pressure spectra were recorded to determine the absorption cross sections of these strong $Q$-branch features. The $Q$-branch cross sections were then added to the high pressure absorption cross sections in place of the high pressure $Q$-branch features, which had been intentionally saturated (see Section~\ref{sect3}).

\begin{table*}[t!]
  \caption{Summary of C$_{2}$H$_{6}$ measurements }
  \label{tab2}
  \centering
  \begin{small}
  \begin{tabular}{@{}ccccccc@{}}
  \rule{0pt}{1ex}  \\
  \hline
  \rule{0pt}{1ex}  \\
        &   Sample     &  Sample C$_{2}$H$_{6}$  &               &             \vspace{-0.20cm}  \\
        &              &                         &    Sample     & Background  \vspace{-0.20cm} \\
  Mode  & Temperature  &   Pressure              &               &             \vspace{-0.20cm} \\
        &              &                         &    Scans      & Scans       \vspace{-0.20cm} \\
        &     (K)      &     (Torr)              &               &             \\
  \rule{0pt}{1ex}  \\
 \hline
  \rule{0pt}{1ex}  \\
  \multirow{10}{*}{Absorption} & 297 & 0.276 & 400 & 550  \\
                               & 297 & 0.035 &  24 &  24  \\
                               & 473 & 0.982 & 300 & 300  \\
                               & 473 & 0.176 &  24 &  24  \\
                               & 573 & 1.476 & 300 & 300  \\
                               & 573 & 0.282 &  24 &  24  \\
                               & 673 & 2.928 & 150 & 150  \\
                               & 673 & 0.987 &  24 &  24  \\
                               & 773 & 5.026 & 150 & 150  \\
                               & 773 & 1.557 &  24 &  24  \\
  \rule{0pt}{1ex}  \\
 \hline
  \rule{0pt}{1ex} \\
  \multirow{4}{*}{Emission}    & 673 & 3.067 & 150 & 150  \\
                               & 673 & 1.008 &  24 &  24  \\
                               & 773 & 5.206 & 150 & 150  \\
                               & 773 & 1.534 &  24 &  24  \\
  \rule{0pt}{1ex} \\
 \hline
  \rule{0pt}{1ex} \\
\end{tabular}
\end{small}
\end{table*}

\section{Absorption cross sections}
\label{sect3}

\begin{figure*}[t!]
  \centering
    \includegraphics[width=0.9\textwidth]{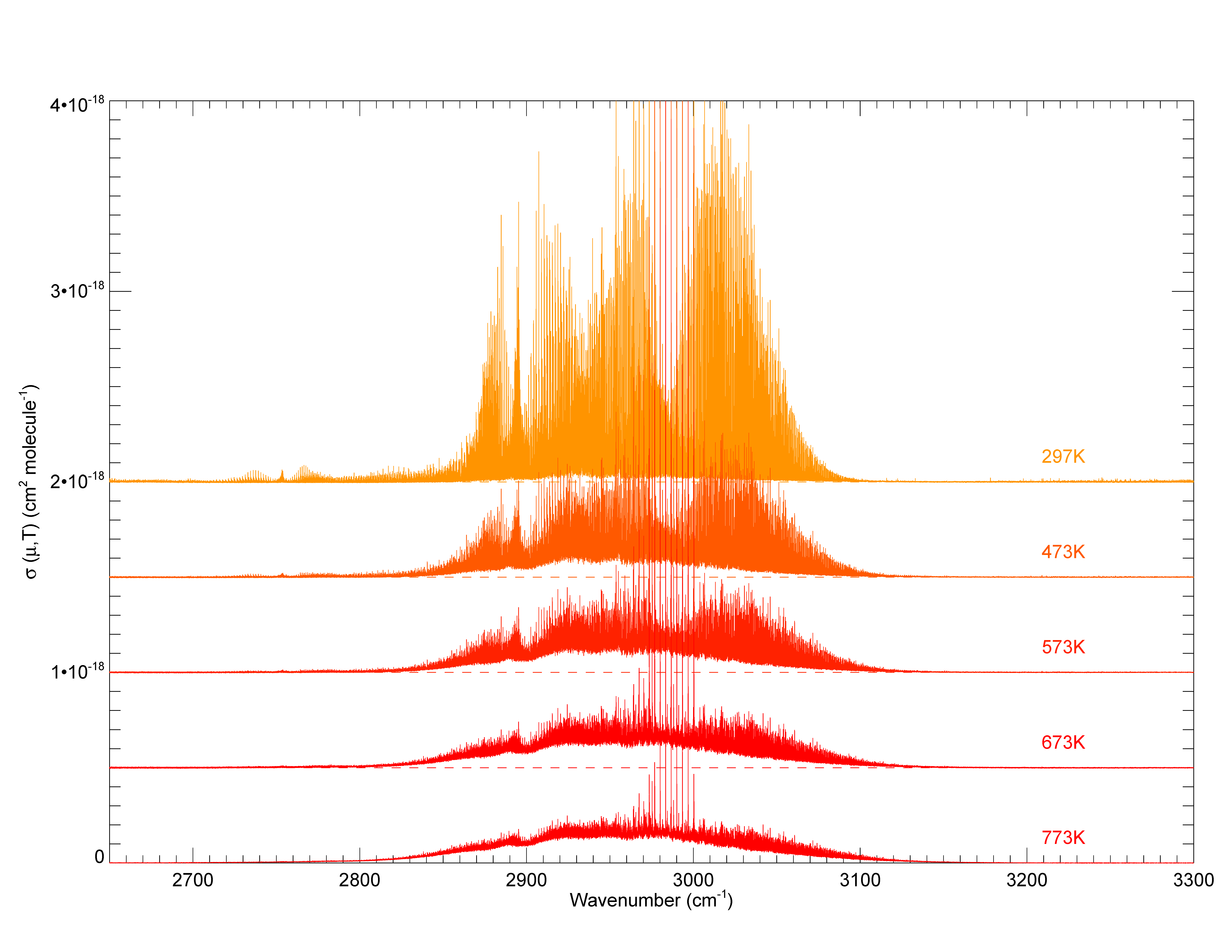}
  \caption{C$_{2}$H$_{6}$ cross sections of the 3.3 $\mu$m region at elevated temperatures. Due to the strength of the $\nu_{7}$ $Q$-branches, each temperature has been offset by 5$\times$10$^{-19}$ cm$^{2}$ molecule$^{-1}$ to highlight the detail in the surrounding region. The baseline for each temperature is given by the corresponding dashed line.}
\label{fig1}
\end{figure*}

\begin{figure*}[t!]
  \centering
    \includegraphics[width=0.9\textwidth]{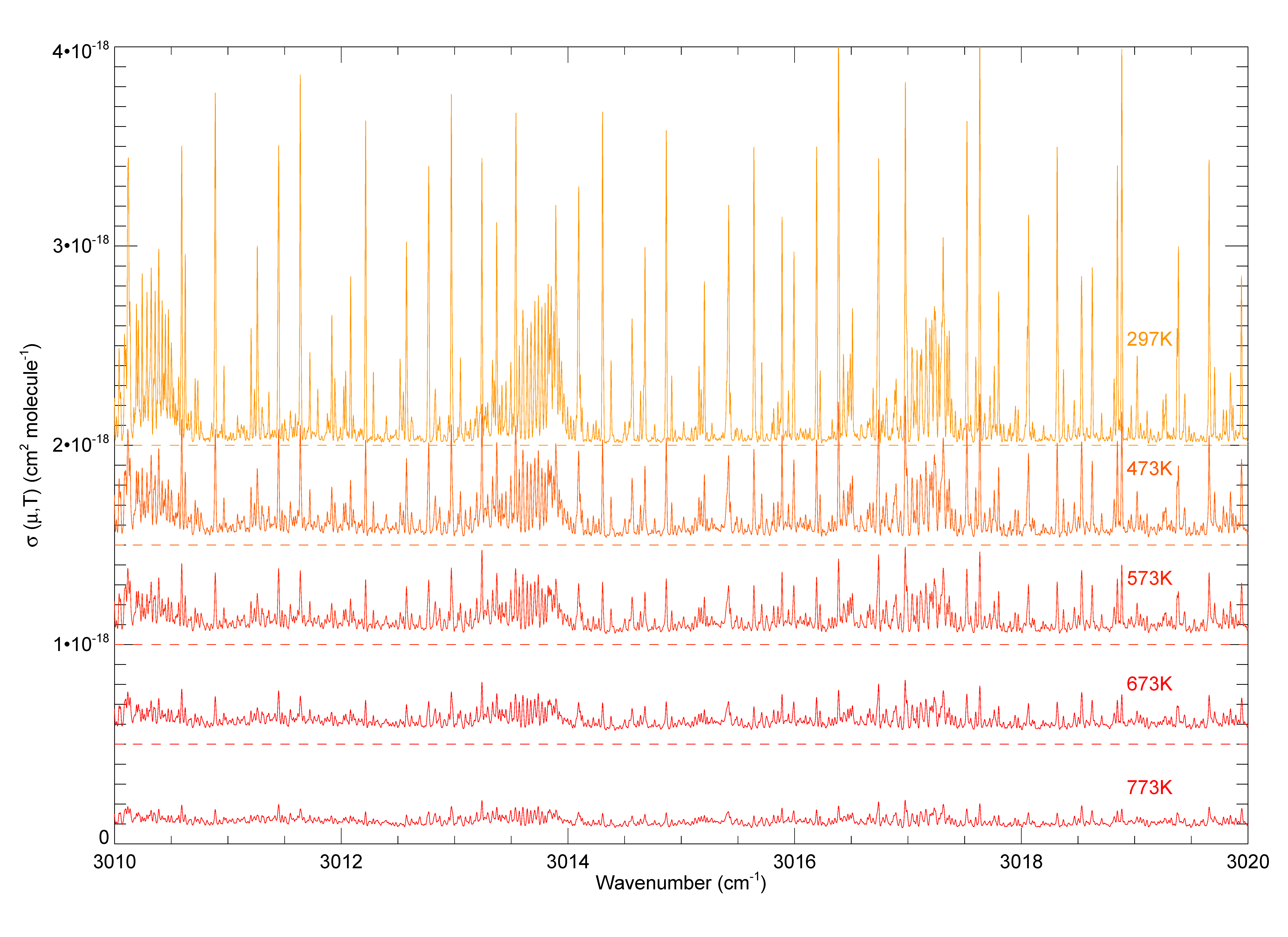}
  \caption{C$_{2}$H$_{6}$ cross sections in the vicinity of the $^{\textrm{r}}$Q$_{7}$ (3010.7 cm$^{-1}$), $^{\textrm{r}}$Q$_{8}$ (3014.0 cm$^{-1}$) and $^{\textrm{r}}$Q$_{9}$ (3017.5 cm$^{-1}$) branches of the $\nu_{7}$ mode. The baseline for each temperature is given by the corresponding dashed line.}
\label{fig2}
\end{figure*}

An absorption cross section, $\sigma$ (cm$^{2}$ molecule$^{-1}$), can be calculated using
\begin{equation}\label{crosssections}
  \sigma = -\xi\frac{10^{4}k_{\textrm{B}}T}{Pl}\ln\tau,
\end{equation}
where $T$ is the temperature (K), $P$ is the pressure of the absorbing gas (Pa), $l$ is the optical pathlength (m),  $\tau$ is the observed transmittance spectrum, $k_{\textrm{\scriptsize{B}}}$ is the Boltzmann constant and $\xi$ is a normalization factor \cite{2010JQSRT.111..357H, 2015JMS..inpressH}.

It has been demonstrated by numerous studies on a variety of molecular spectra that integrating an absorption cross section over an isolated band (containing primarily fundamentals) exhibits an insignificant temperature dependence \cite{1958JChPh..29.1042C,1959JChPh..30.1619M,1965JChPh..42..402B,1976AcSpA..32.1059Y,2010JQSRT.111.1282H,2010JQSRT.111..357H,2012JQSRT.113.2189H}.

The PNNL infrared absorption cross sections of C$_{2}$H$_{6}$ cover the spectral range 600--6500 cm$^{-1}$ (resolution of 0.112 cm$^{-1}$) at 278, 293 and 323 K. Each PNNL cross section is a composite of approximately ten pathlength concentrations, making these data suitably accurate for calibration \cite{2010JQSRT.111..357H}. For the spectral region between 2500 and 3500 cm$^{-1}$ the average PNNL integrated absorption is calculated as
\begin{equation}\label{intint}
\begin{aligned}
\begin{split}
  \int^{3500 \textrm{ cm}^{-1}}_{2500 \textrm{ cm}^{-1}} \sigma(\nu,T) d\nu  = & 2.976(\pm0.011) \times 10^{-17} \\ & \textrm{ cm molecule}^{-1}.
\end{split}
\end{aligned}
\end{equation}
Each individual PNNL cross section demonstrates less than 0.4\% deviation from this value\footnote{PNNL units (ppm$^{-1}$m$^{-1}$ at 296 K) have been converted using the factor $k_{B}\times296\times\ln(10)\times10^{4}/0.101325 = 9.28697\times10^{-16}$ \cite{2012JQSRT.113.2189H}}.

The new transmittance spectra have been converted into cross sections using Equation~\ref{crosssections}, making the original assumption that $\xi=1$. This is to allow the strong $\nu_{7}$ $Q$-branch features from the low pressure observations to be inserted in place of the same saturated (therefore distorted) $Q$-branch features in the high pressure absorption cross sections. Each replaced $Q$-branch region covered less than $\sim$0.2 cm$^{-1}$ and was chosen to be between the points where the high and low pressure cross sections {\color{Black}intersect} either side of the strong feature. These composite absorption cross sections were then integrated over the 2500 and 3500 cm$^{-1}$ spectral region. Comparisons were made to the PNNL integrated absorption cross section (Equation~\ref{intint}) in order to calibrate our observations. The normalization factors for each absorption cross section are provided in Table~\ref{tab3}, alongside the calibrated pressures and calibrated integrated absorption cross sections.

\begin{table*}[t!]
  \caption{Parameters for calibrated absorption cross sections}
  \label{tab3}
  \centering
  \begin{small}
  \begin{tabular}{@{}ccccccc@{}}
  \rule{0pt}{1ex}  \\
  \hline
  \rule{0pt}{1ex}  \\
              &   Normalization &  C$_{2}$H$_{6}$  &  Integrated absorption  \vspace{-0.20cm} \\
 Temperature  &                 &                  &                         \vspace{-0.20cm} \\
              &      Factor     &    Effective     &      cross section      \vspace{-0.20cm} \\
     (K)      &                 &                  &                         \vspace{-0.20cm} \\
              &      $\xi$      & Pressure (Torr)  &  ($\times10^{-17}$ cm molecule$^{-1}$)   \\
  \rule{0pt}{1ex}  \\
 \hline
  \rule{0pt}{1ex}     \\
 297 & 1.028 & 0.269 & 2.978  \\
 473 & 1.035 & 0.948 & 2.978  \\
 573 & 1.063 & 1.388 & 2.976  \\
 673 & 1.047 & 2.796 & 2.975  \\
 773 & 1.033 & 4.867 & 2.976  \\
  \rule{0pt}{1ex} \\
 \hline
  \rule{0pt}{1ex} \\
\end{tabular}
\end{small}
\end{table*}

The calibrated cross sections are displayed in Figure~\ref{fig1} between 2600 and 3300 cm$^{-1}$ and clearly display the $\nu_{5}$ and $\nu_{7}$ {\color{Black}fundamental bands}. Figure~\ref{fig2} shows a 10 cm$^{-1}$ section of Figure~\ref{fig1} in the vicinity of three weak $\nu_{7}$ $Q$-branch features; an increase in the observed continuum at higher temperatures can be seen.

\begin{figure*}[t!]
  \centering
    \includegraphics[width=0.75\textwidth]{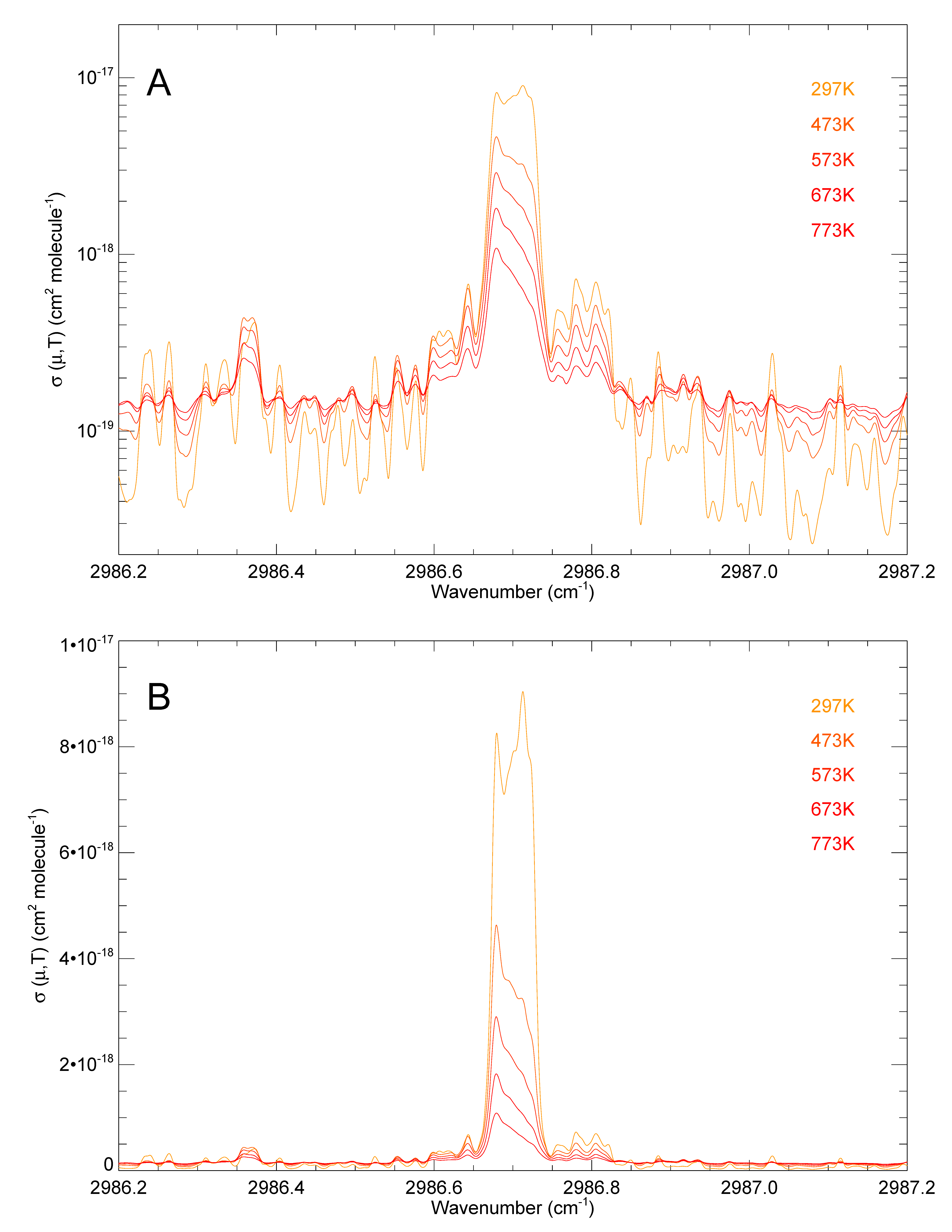}
  \caption{Temperature dependance of the $^{\textrm{r}}$Q$_{0}$-branch of the $\nu_{7}$ mode of C$_{2}$H$_{6}$. (A) The decreasing $^{\textrm{r}}$Q$_{0}$-branch is observed as the surrounding continuum increases, using a logarithmic scale. (B) The shape of the $^{\textrm{r}}$Q$_{0}$-branch is seen to change shape at higher temperatures on a linear scale. }
\label{fig3}
\end{figure*}

The calibrated absorption cross sections described in Table~\ref{tab3} are available online\footnote{http://bernath.uwaterloo.ca/C2H6/} in the standard HITRAN format \cite{2013JQSRT.130....4R}.

\section{Discussion}
\label{sect4}

The normalization factors are necessary to account for the difficulty in measuring the experimental parameters accurately (i.e., pathlength, pressure and temperature). The combination of the errors often leads to an underestimation of the integrated absorption cross section, which is calibrated  by comparison to the PNNL data. Normalization factors are typically within 6\% for measurements using similar apparatus \citep[e.g.,][]{2010JQSRT.111..357H}. For our measurements, the normalization factor has been used to give an effective calibrated pressure as seen in Table~\ref{fig3}. Based upon consideration of the experimental and photometric errors, the calibrated cross sections are expected to be accurate to within 4\%. 

{\color{Black}The C$_{2}$H$_{6}$ absorption cross sections available at 194 K are also based on a calibration to the PNNL \cite{2010JQSRT.111..357H}}. These data contains C$_{2}$H$_{6}$ at 0.2208 Torr, which has been broadened by 103.86 Torr of air at a resolution of 0.015 cm$^{-1}$. Integrating the available data between 2545 and 3315 cm$^{-1}$  yields a value of 2.985$\times10^{-17}$ cm molecule$^{-1}$. This is within 0.3\% of the average values contained in Table~\ref{tab3}. An independent quality check can be made by comparing to new C$_{2}$H$_{6}$ absorption cross sections recorded for combustion applications \cite{2014JMoSp.303....8A}. These data contain medium resolution (0.16-0.6 cm$^{-1}$) nitrogen-broadened cross sections of C$_{2}$H$_{6}$ between 2500 and 3400 cm$^{-1}$. At temperatures of 296, 673 and 773 K the C$_{2}$H$_{6}$ {\color{Black}integrated} absorption cross sections were calculated to be 2.81$\times10^{-17}$, 2.99$\times10^{-17}$ and 3.08$\times10^{-17}$ cm molecule$^{-1}$, respectively. While a small temperature dependence is seen when compared to our values in Table~\ref{tab3}, the deviation is within their experimental error (5\%).

Experimental spectra of C$_{2}$H$_{6}$ have not been acquired at temperatures above 773 K as the molecules begin to decompose when using a sealed cell. Evidence of CH$_{4}$ absorption was observed at 873 K; therefore reliable C$_{2}$H$_{6}$ cross sections could not be obtained.

C$_{2}$H$_{6}$ is expected to be useful as a temperature probe for exoplanets and brown dwarfs \cite{2013A&ARv..21...63T}. The infrared spectrum (Figures~\ref{fig1} and~\ref{fig2}) demonstrates a continuum-like feature previously observed for CH$_{4}$ at high temperatures \cite{2015ApJ..inpressH}. It can be seen that as the continuum increases with temperature, the sharp $Q$-branches decrease due to a change in the population of states and they also broaden because of the increasing Doppler width (Figure~\ref{fig3}). However, Table~\ref{tab3} shows $\xi$ only exhibits a small change and the integrated intensity remains constant (within experimental error). This variation is small enough to justify the assumption that the integrated absorption cross sections are independent of temperature. 

For weak concentrations of C$_{2}$H$_{6}$ it may be difficult to observe a change in the continuum, particularly since the 3.3 $\mu$m region contains the prominent C-H stretch for hydrocarbons. However, the shape of the sharp $Q$-branches of the $\nu_{7}$ mode are also seen to change with increasing temperature, as shown in Figure~\ref{fig3}. These $Q$-branches are relatively easy to identify in congested atmospheric spectra, therefore studying the shape of these features can also be used to infer temperatures. 

\section{Conclusion}
\label{sect5}

High-resolution infrared absorption cross sections for C$_{2}$H$_{6}$ have been measured at elevated temperatures (up to 773 K) between 2500 and 3500 cm$^{-1}$. The spectra were recorded at a resolution of 0.005 cm$^{-1}$ and the integrated absorption has been calibrated to PNNL values. These data are of particular interest for simulating astronomical environments at elevated temperatures, such as brown dwarfs and exoplanet atmospheres, where C$_{2}$H$_{6}$ can be used as a temperature probe. With the imminent arrival of the Juno spacecraft into orbit around Jupiter, these data will be of particular use for observations made of aurora using the JIRAM instrument.

\paragraph{Acknowledgments}

Funding was provided by the NASA Outer Planets Research Program.

\section*{References}

\bibliography{Hot-Ethane-BIBFILE}

\end{document}